\documentclass[prl,aps,floats,twocolumn,showpacs]{revtex4}
\usepackage[dvips]{epsfig}

\begin{document}

\title{Finite Volume Kolmogorov-Johnson-Mehl-Avrami Theory}

\author{Bernd A. Berg$^{\rm \,a,b,}$\footnote{Corresponding author.}
and Santosh Dubey$^{\rm \,a,b}$ }

\affiliation{$^{\rm \,a)}$ Department of Physics, Florida State
University, Tallahassee, FL 32306-4350, USA\\
$^{\rm \,b)}$ School of Computational Science,
Florida State University, Tallahassee, FL 32306-4120, USA}

\date{February 4, 2008; revised March 17, 2008.} 

\begin{abstract}
We study Kolmogorov-Johnson-Mehl-Avrami (KJMA) theory of phase 
conversion in finite volumes. For the conversion time we find the
relationship $\tau_{\rm con} = \tau_{\rm nu}\,[1+f_d(q)]$. Here $d$ 
is the space dimension, $\tau_{\rm nu}$ the nucleation time in the 
volume $V$, and $f_d(q)$ a scaling function. Its dimensionless argument 
is $q=\tau_{\rm ex}/\tau_{\rm nu}$, where $\tau_{\rm ex}$ is an 
expansion time, defined to be proportional to the diameter of the 
volume divided by expansion speed. We calculate $f_d(q)$ in one, two 
and three dimensions. The often considered limits of phase conversion 
via either nucleation or spinodal decomposition are found to be 
volume-size dependent concepts, governed by simple power laws for 
$f_d(q)$.
\end{abstract}
\maketitle


Phase conversions are of importance in physics, chemistry and other 
fields.  Examples are numerous and include crystal physics \cite{Zh65}, 
metallurgy \cite{Sm07}, polymer physics \cite{Bi95,St07},
ferroelectric domain switching \cite{IT70}, magnetization and
metastability in statistical physics models \cite{RG94,LB00}, 
phase transitions in particle physics \cite{MO07}, as well as 
ecological landscapes~\cite{OM06}.

Specific phenomena are nucleation and spinodal decomposition
\cite{spinodal}.  Conventionally, for a review see \cite{La92}, 
nucleation is characterized by metastability, while spinodal 
decomposition is considered to be the mechanism by which phase 
conversion occurs in an unstable system. We shall discuss the 
crossover of these phenomena as function of the nucleation time, 
the expansion speed, and the volume.

In Kolmogorov-Johnson-Mehl-Avrami (KJMA) theory \cite{Ko37,JM39,Av39}
phase conversion is based on the rate of nucleation of critical 
nuclei \cite{droplets} and their subsequent expansion speed due to
gain in free energy. Independently, this approach was developed a few 
years later in a concise paper by Evans \cite{Ev45}. More recent work 
derived space-time correlation functions \cite{Se90} and dealt with 
screening effects \cite{Bu05}. KJMA theory is formulated in an infinite 
volume. But in physics there are no truly infinite volumes. Our 
investigation of finite volumes leads to interesting scaling laws.


There are several scenarios of KJMA theory. Some deal also with 
non-critical nuclei \cite{Av39}.
We consider here the limit in which critical nuclei, compared 
to the total volume are small enough to be considered pointlike. 
Extension of our considerations are possible, but would at present 
distract from the main point. 

In accordance with KJMA theory we make the following assumptions:
\medskip

\begin{enumerate}

\item Critical nuclei are created with a constant nucleation rate
      $R=1/(\tau_{\rm nu}V)$ at uniformly distributed random positions 
      in the volume $V$. Let us denote by $V^0$ the unit volume, and 
      by $\tau^0_{\rm nu}$ the nucleation time (average time it
      takes to create a critical nucleus in the unit volume). Then
      the nucleation time in the volume $V$ is given by $\tau_{\rm nu}
      =\tau^0_{\rm nu} V^0/V$.

\item Subsequent growth: A nucleus created at time $t_i$ covers at 
      time $t>t_i$ the spherical volume $V_i(t)=C_d\,[v\,(t-t_i)]^d$,
      where $v$ is the expansion speed, $d$ the space dimension  
      and $C_d$ a dimension dependent factor ($C_1=2$, $C_2=\pi$,
      and $C_3=4\pi/3$).

\item The converted volume $V_{\rm con}(t)$ is the union of the
      volumes $V_i(t)$ ($V_i(t)=0$ for $t\le t_i$), intersected 
      by the total volume $V$.

\end{enumerate}

Assumption~1 allows the creation of nuclei in the already converted 
volume $V_{\rm con}(t)$. From assumptions~2 and~3 it is clear that 
they do not contribute to phase conversion and, therefore, they are 
not added to the number of nuclei in the volume $V$. Note that KJMA 
theory of phase conversion is kinetic with no details of the 
responsible interactions involved.

The time it takes to transform the bulk system into the new phase is 
the conversion time $\tau_{\rm con}$. There is some arbitrariness in
its definition. In essence any converted volume in the range $0.5\le 
V_{\rm con}(\tau_{\rm con})/V<1$ , e.g. $V_{\rm con}(\tau_{\rm con})
/V=0.90$, defines a suitable conversion time. Only, $V_{\rm con}
(\tau_{\rm con}) /V=1$ is not admissible: $\tau_{\rm con}$ will then 
diverge in the infinite volume limit, because due to statistical 
fluctuations some points stay always unconverted in an infinite volume. 
This is well-known in KJMA theory and even more obvious for systems 
with fluctuations due to interactions. 

For practical reasons we define the conversion time by distributing 
a finite number of trial points uniformly over the volume and its 
boundaries and taking $\tau_{\rm con}$ as the time at which all 
points are first covered by the new phase. The number of points is 
taken to be a constant, independent of the size of the volume. We 
restrict ourselves to cubic volumes of size $V=L^d$, and choose as trial 
points the sites of a hypercubic lattice that includes the $2^d$ corner 
points of $V$.  Extensions to other geometries are straightforward. In 
particular geometries can be chosen to fit actual experimental situations.

To calculate the average conversion time turns out to be easier than 
one might expect. There are only two independent parameters with the 
dimension of a time, $\tau^0_{\rm nu}$ and an expansion time 
$\tau_{\rm ex}$, which we define by  
\begin{equation} \label{tauex}
  \tau_{\rm ex} =\frac{L}{v}\ . 
\end{equation}
The functional dependence $\tau_{\rm con}(\tau^0_{\rm nu},v,V)$ is 
determined by a scaling function $f_d(q)$ \cite{Be07} as presented 
below. Instead of $\tau^0_{\rm nu}$ we use the nucleation time 
$\tau_{\rm nu}$ of the total volume $V$,
\begin{equation} \label{taunu}
  \tau_{\rm nu} = \frac{\tau^0_{\rm nu}}{\lambda^d}~~{\rm with}~~
  \lambda^d = \frac{V}{V^0}\ ,
\end{equation}
and the scaling function $f_d(q)$ is defined by
\begin{equation} \label{taucon}
  \frac{\tau_{\rm con}}{\tau_{\rm nu}} =\left[ 1 + f_d(q)\right]~~
  {\rm with}~~ q = \frac{\tau_{\rm ex}}{\tau_{\rm nu}}\ .
\end{equation}
The reduction from three variables to one is a mayor simplification.
That $\tau_{\rm con}/\tau_{\rm nu}$ depends indeed only on $q$ is shown 
in the following. Natural independent variables are $\tau^0_{\rm nu}$,
$v$ and $L$. Using $\tau_{\rm nu}$ instead of $\tau^0_{\rm nu}$ as 
independent variable is mathematically equivalent. One can then 
define three transformations, which leave $q$ invariant: 
(1)~$L_1'=\lambda_1 L$, $v_1'=\lambda_1 v$; 
(2)~$L_2'=\lambda_2 L$, $\tau'_{2,\rm nu}=\lambda_2 \tau'_{\rm nu}$;
(3)~$v'_3=\lambda_3 v$, $\tau'_{3,\rm nu}=\tau'_{\rm nu}/\lambda_3$.
Combinations of the three transformations allow us to create all values
$\tau'_{\rm nu}$, $v'$ and $L'$ for which $q'=q$ holds. Now, nuclei
at the appropriately scaled positions in the volume $V'=(L')^d$ are
created with the same probabilities as in the initial volume $V$. 
In case~(1) the change in volume (length) is compensated by the
increase in velocity, so that the conversion time stays constant,
$\tau'_{1,\rm con}=\tau_{\rm con}$. In case~(2) the nucleation
time is scaled, so that the fixed velocity $v$ creates up to scaling
in size the same patterns as before. Therefore, the conversion time 
scales according to $\tau'_{2,\rm con}=\lambda_2 \tau'_{\rm con}$ and 
$\tau'_{2,\rm con}/\tau'_{2,\rm nu}=\tau_{\rm con}/\tau_{\rm nu}$.
Similarly, in case~(3) the change in velocity is compensated by
the change of the nucleation time, so that the created patterns 
stay the same and $\tau'_{3,\rm con}/\tau'_{3,\rm nu}=
\tau_{\rm con}/\tau_{\rm nu}$ holds.

In the limit of large expansion speeds ($v\to\infty$, volume fixed) 
we find
\begin{equation} \label{qtozero}
  f_d^{\rm small}(q)=A_d\,q~~{\rm for}~~ q\to 0\ ,
\end{equation}
where $A_d$ is a dimension and geometry dependent constant. In this 
limit, creation of a first critical nucleus takes much longer than its 
subsequent expansion to the size of the volume $V$. Therefore, creation 
of several critical nuclei is unlikely and $\tau_{\rm nu}$ becomes the 
time of metastability. The conversion time is determined by the farthest 
away corner of the $L^d$ volume, once the nucleus is created. By 
integration over the possible positions of the nucleus one finds 
$A_1=0.75$, $A_2=1.0704$, $A_3=1.315$, and (for string theorists) 
$A_{10}=2.4110$. The limit (\ref{qtozero}) describes the nucleation 
scenario of phase conversion. 

At large $q$, the function $f_d(q)$ is up to a multiplicative constant 
also analytically determined. Imagine, we calculate the conversion 
time simultaneously on $n^d$ non-interacting systems with identical 
parameters (nucleation time, volume, expansion speed). The conversion 
time is a random variable, which has the same mean value $\tau_{\rm 
con}$ on each system. Let us combine them into one. For $\tau_{\rm ex}
\gg \tau_{\rm nu}$ the effects due to propagation of phase conversion
over the boundaries becomes negligible and $\tau'_{\rm con}$ averaged 
over the combined system is $\tau'_{\rm con}=\tau_{\rm con}$. As we 
have $q\to q'=n^{d+1}q$ and $\tau_{\rm nu}\to\tau'_{\rm nu} = n^{-d}
\tau_{\rm nu}$ for $V\to V' =n^dV$, invariance of the conversion time 
requires
\begin{equation} \label{qtoinfty}
  f_d^{\rm large}(q)=B_d\,q^{d/(d+1)}~~{\rm for}~~ q\to \infty\ .
\end{equation}
This is the limit of spinodal conversion, obtained in volumes of 
fixed size for small expansion speeds, $v\to 0$. Many critical nuclei 
contribute then to the phase conversion. For physical parameters 
$\tau^0_{\rm nu}$, $v$ fixed, and volume $V\to\infty$, i.e., 
$\lambda\to\infty$ in Eq.~(\ref{taunu}), the theory always describes
spinodal decomposition, because $q$ scales as $q\to\lambda^{d+1}\,q$. 

This is in contradiction to the mean-field approach, which leads on 
infinite volumes to a nucleation region with a so called spinodal 
endpoint \cite{LB00,La92}. Within the more realistic scenario of 
KJMA theory the spinodal can only be an effective concept for finite 
volumes. In contrast to the comparison with mean-field theory, our
results are consistent with studies of magnetic field driven phase
conversion by Rikvold et al.~\cite{Ri94}, in which a ``dynamical
spinodal field'' separates the two regimes.

Let us turn to the general evaluation of $f_d(q)$ by Monte Carlo 
(MC) simulations (here not Markov chain MC). The implementation of 
the nucleation process is relatively straightforward and allows 
variations of the expansion speed, and hence $q$, over many orders 
of magnitudes. This comes, because we have to implement only kinetics 
and no complicating dynamics (for instance, due to interactions 
between spins). We use 100 trial points in 1D, $10\times 10 =100$ 
in 2D, and $5\times 5\times 5 =125$ in 3D. For a volume of edge 
length one this corresponds, in the lattice of trial points, to a 
lattice spacing of 1/99 in 1D, 1/9 in 2D, and 1/4 in 3D.

\begin{figure}[tb] \begin{center}
\epsfig{figure=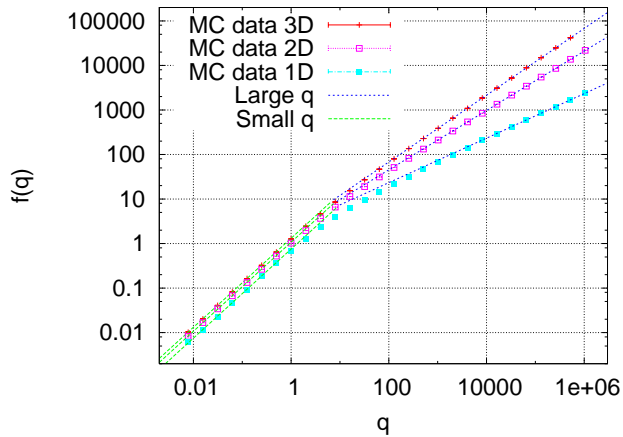,width=\columnwidth} 
\caption{Scaling function $f_d(q)$ versus $q$.
\label{fig_q} } \end{center} \end{figure} 

\begin{table}[tb]
\caption{$B_d$ (\ref{qtoinfty}) for our trial points from MC 
simulations.} \label{tab_KJMA}
\medskip \centering
\begin{tabular}{|c|c|c|c|}   \hline
                       & 1D         & 2D         & 3D         \\ \hline
 $q_d^{\min}$          & 0.5        &   1        &  1         \\ \hline
 $q_d^{\max}$          & 4$\,$096   & 128        & 64         \\ \hline
 $B_d$                 & 2.3285 (60)& 2.0990 (53)& 2.1427 (43)\\ \hline
$n_d$, $\chi^2_d({\rm pdf})$& 6, 1.24& 7, 1.07   & 10, 0.74   \\ \hline
\end{tabular} \end{table}

The results in 1D, 2D and 3D together with the analytical $q\to 0$ and 
$q\to\infty$ asymptotic behavior are presented in Fig.~\ref{fig_q} on 
a log-log scale. When taking data our stepsize was a factor two in $q$.
For crosschecks at a few $q$ values various combinations of $\tau^0_{
\rm nu}$, $v$ and $V$ were used that combine to the same $q$ value.

Performing Gaussian difference tests (e.g., Ref.~\cite{Bbook}), the 
first four data are in each case consistent with the small $q$ 
approximation (\ref{qtozero}). For $q\le q_d^{\min}$, $q_d^{\min}$ 
listed in table~\ref{tab_KJMA}, the data are found to agree with an 
relative error $|f_d(q)-f_d^{\rm small}(q)|/f_d(q)<5\%$ with the
analytical small $q$ behavior. In the same way they are consistent 
with the large $q$ behavior (\ref{qtoinfty}) for $q\ge q_d^{\max}$, 
where the $B_d$ values listed in table~\ref{tab_KJMA} are determined 
by one-parameter fits to $n_d$ data with the largest $q$ values 
(chi-squared per degree of freedom of the fit, $\chi^2_d({\rm pdf})=
\chi^2_d/(n_d-1)$, is also given). In the sense of these approximations 
we have nucleation for $q<q_d^{\min}$, spinodal decomposition for 
$q>q_d^{\max}$, and a crossover region in-between. 

In particular, we have in this classification for a 3D cubic box
$\tau_{\rm nu}/\tau_{\rm con}<0.02$ for spinodal decomposition and 
$\tau_{\rm nu}/\tau_{\rm con}>0.43$ for nucleation. While the 
coefficients $A_d$ and $B_d$ in Eqs.~(\ref{qtozero}) 
and~(\ref{qtoinfty}) depend on the geometry, the power laws do not. 
Therefore, they can be employed to characterize the limits universally.

\begin{figure}[tb] \begin{center}
\epsfig{figure=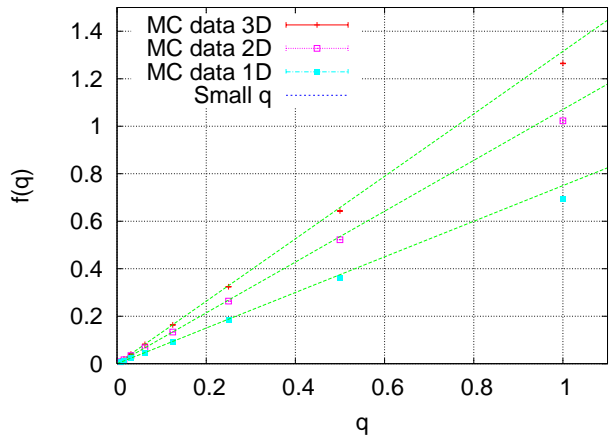,width=\columnwidth} 
\caption{Scaling function: End of the small $q$ region~(\ref{qtozero}). 
\label{fig_qsmall} } 
\end{center} \end{figure} 

\begin{figure}[tb] \begin{center}
\epsfig{figure=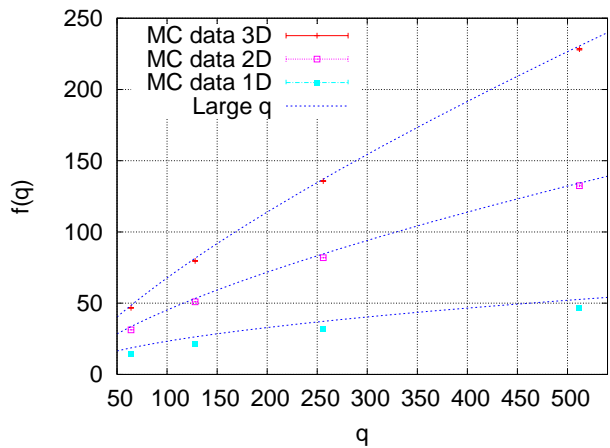,width=\columnwidth} 
\caption{Scaling function: Beginning of the large $q$ 
region~(\ref{qtoinfty}) in 2D and 3D. \label{fig_qlarge} } 
\end{center} \end{figure} 

In Fig.~\ref{fig_qsmall} we exhibit the end of the small $q$ region 
on a scale with higher resolution than that of the logarithmic scale 
of Fig.~\ref{fig_q}. Correspondingly the beginning of the large $q$ 
region in 2D and 3D is shown in Fig.~\ref{fig_qlarge}. In 1D the 
asymptotic behavior sets in for considerably larger $q$ values. The 
reason appears to be that there are not yet sufficiently many nuclei 
participating in the phase conversion. At the largest $q$ value of 
Fig.~\ref{fig_qlarge} $(q=512)$ the deviation of the 1D MC result from 
its large $q$ asymptotics is still 11\%.

\begin{figure}[tb] \begin{center}
\epsfig{figure=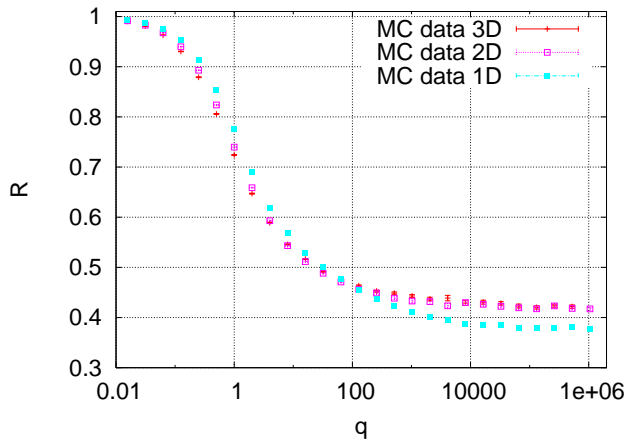,width=\columnwidth} 
\caption{Ratio of contributing nuclei versus $q$. \label{fig_nuclei} } 
\end{center} \end{figure} 

The average number of participating nuclei, $N_{\rm nuclei}$ is smaller 
than $\tau_{\rm con}/\tau_{\rm nu}$ of Eq.~(\ref{taucon}) as nuclei
created inside an already converted region do not contribute. In 
Fig.~\ref{fig_nuclei} we show the ratio $R=N_{\rm nuclei}/(
\tau_{\rm con}/\tau_{\rm nu})$. For large $q$ it approaches $R=0.38$ 
in 1D and $R=0.42$ in 2D and 3D. These numbers are specific to our
choice of trial points.


We continue with illustrations.
Changes in physical conditions, for instance of the temperature, can
influence the nucleation time $\tau^0_{\rm nu}$, the expansion speed
$v$, and the volume $V$. 

Let us assume a constant volume. If the nucleation time varies from 
$\tau_{\rm nu}\to \tau'_{\rm nu}$ for fixed expansion velocity $v$, 
while we stay in the spinodal region $q>q_d^{\max}$, the scaling 
$q\to q'=(\tau_{\rm nu}/\tau'_{\rm nu})\,q$ yields for the conversion 
time change $\tau_{\rm con}' = (\tau'_{\rm nu}/\tau_{\rm nu})^{1/(d+1)} 
\,\tau_{\rm con}$. If in the same situation the nucleation time 
$\tau_{\rm nu}$ is fixed and the expansion speed varies from $v\to v'$, 
we find for the new conversion time $\tau_{\rm con}'=(v/v')^{d/(d+1)}\,
\tau_{\rm con}$.  When staying in the nucleation region the 
corresponding equations are $\tau'_{\rm con} = (\tau'_{\rm nu}/
\tau_{\rm nu})\,\tau_{\rm con}+A_d\,(\tau_{\rm nu}-\tau'_{\rm nu})\,q$ 
and $\tau'_{\rm con}/\tau_{\rm nu}=1+A_d\,(v/v')\,q$, respectively.

Assume a 2D Ising model on a $100\times 100$ lattice is prepared in 
its initial state with all spins down. It is then simulated by Markov
chain Monte Carlo \cite{Bbook} below the critical temperature and with a 
magnetic field opposite to the initial orientation of the spins. 
For suitable choices of temperature and magnetic field the following 
numbers are realistic: (A)~Seven nucleation events in one sweep with 
a subsequent expansion speed of 5 lattice spacings in 20 sweeps. 
(B)~One nucleation event in 1680 sweeps and a subsequent expansion 
speed of 50 lattice site in 800 sweeps. A brief calculation puts 
case~(A) with $q=2800$ solidly into the spinodal asymptotics, while 
with $q=0.95$ case~(B) is at the end of nucleation region. Enlarging 
the lattice to $1\,000\times 1\,000$ sites moves case~(B) to $q=953$ 
into the the spinodal region.

Consider a metastable liquid in a cubic box of size $(0.1\,{\rm meter}
)^3$ with a nucleation time of 1 minute in that volume and a subsequent 
explosion-like conversion at the speed of 100 km/h. With $q=6\times 
10^{-5}$ this is deep in the nucleation region. This is no longer true
if the same system is a pool of size $(10\,{\rm meter})^3$. Then we 
are at $q=6\,000$, though the preparation of such a large homogeneous 
system may in practice be impossible.

Conversion times of the order of minutes are observed in polyethylene
crystallization \cite{Lu01}. To be definite, let $\tau_{\rm con}=180
\,$s. If the nucleation time for the relevant volume is $\tau_{\rm nu}
\le 3.6\,$s, we would classify the process as spinodal decomposition, 
and for $\tau_{\rm nu}\ge 77.8\,$s as nucleation, with the crossover 
region in the range $3.6\,{\rm s}<\tau_{\rm nu}<77.8\,$s.

Let us consider the deconfining phase transition \cite{MO07} and choose 
$(1\,{\rm fermi})^3$ as the unit volume which defines $\tau^0_{\rm nu}$. 
Suppose the relevant volume at 
a heavy ion collider is of size $(10\,{\rm fermi})^3$, and that the 
deconfined phase spreads out at the speed of light once a nucleus is 
created. What is the range of nucleation times so that the phase
conversion (confined $\to$ deconfined) proceeds by spinodal 
decomposition ($q\ge 64$)? The answer is $\tau^0_{\rm nu}< 5\times
10^{-22}$ seconds. This estimate goes up when the expansion speed
is slower than the speed of light.


Conclusions.
Our equations will need corrections, once the critical nuclei can
no longer be considered pointlike, and their size introduces a new 
dimensional parameter. Further, correlations between nuclei are 
presently neglected and the constant expansion speed of KJMA theory
may be a too crude approximation for the actual dynamical, stochastic
expansion process. Nevertheless, we think that the scaling laws 
outlined in this paper are at the heart of the distinction between
nucleation and spinodal regimes of phase transitions.

\acknowledgments 
We like to thank Per Arne Rikvold and Sachin Shanbhag for useful 
discussions. This work was in part supported by the DOE grant 
DE-FG02-97ER41022.


\begin{thebibliography}{19}
\bibitem{spinodal} Here we use the terminology ``spinodal decomposition''
                   and ``spinodal conversion'' in a broader sense than 
                   some statistical physicists do.
\bibitem{droplets} The nuclei are also called droplets, for instance
                   in Ref.~\cite{RG94}.
\end{thebibliography}
\end{document}